%% file: main.tex
\documentclass[a4paper,twoside]{article}

\usepackage{graphicx}
\usepackage{subcaption}
\usepackage{calc}
\usepackage{amssymb}
\usepackage{amstext}
\usepackage{amsmath}
\usepackage{amsthm}
\usepackage{multicol}
\usepackage{pslatex}
\usepackage{apalike}
\usepackage{amsmath}
\usepackage{algorithm, listings}
\usepackage{verbatim}
\usepackage[noend]{algpseudocode}
\algrenewcommand{\algorithmicrequire}{\textbf{Input:}}
\algrenewcommand{\algorithmicensure}{\textbf{Output:}}
\algnewcommand\And{\textbf{and} }\usepackage{amsthm}
\usepackage{hyperref}
             \usepackage{acro}
           
\usepackage{SCITEPRESS}     

\DeclareAcronym{scn}{
      short = SCN ,
      long  = Supply Chain Network ,
      tag = abbrev
    }
    \DeclareAcronym{kpi}{
      short = KPI ,
      long  = Key Performance Indicator ,
      tag = abbrev
    }

   \DeclareAcronym{sc}{
      short = SC ,
      long  = Supply Chain ,
      tag = abbrev
    }
       \DeclareAcronym{kg}{
      short = KG ,
      long  = Knowledge Graph ,
      tag = abbrev
    }

\DeclareAcronym{dmp}{
      short = DMP ,
      long  = Disruption Management Process ,
      tag = abbrev
    }

\begin{document}

\title{MARE: Semantic Supply Chain Disruption\\ Management and Resilience Evaluation Framework}

\author{\authorname{Nour Ramzy \sup{1}
, Sören Auer \sup{2}
, Hans Ehm \sup{1}
, and Javad Chamanara \sup{2}
}
\affiliation{\sup{1}Infineon Technologies AG, Am Campeon 1-15, 85579 Munich, Germany }
\affiliation{\sup{2}TIB: Technische Informationsbibliothek Welfengarten 1B, 30167 Hannover, Germany}
\email{\{nour.ramzy, hans.ehm\}@infineon.com, \{soeren.auer, javad.chamanara\}@tib.eu}
}

\abstract{ \acp{sc} are subject to disruptive events that potentially hinder the operational performance.
 \ac{dmp} relies on the analysis of integrated heterogeneous  data sources such as production scheduling, order management and logistics to evaluate the impact of disruptions on the \ac{sc}. 
Existing approaches are limited as they address \ac{dmp} process steps and corresponding data sources in a rather isolated manner which hurdles the systematic handling of a disruption originating anywhere in the \ac{sc}. 
Thus, we propose MARE a semantic disruption management and resilience evaluation framework for integration of data sources included in all \ac{dmp} steps, i.e. \underline{M}onitor/Model, \underline{A}ssess, \underline{R}ecover and \underline{E}valuate. 
MARE, leverages semantic technologies i.e. ontologies, knowledge graphs and SPARQL queries to model and reproduce \ac{sc} behavior under disruptive scenarios.
Also, MARE includes an evaluation framework to examine the restoration performance of a \ac{sc} applying various recovery strategies. 
Semantic \ac{sc} \ac{dmp}, put forward by MARE, allows stakeholders to potentially identify the measures to  enhance \ac{sc} integration, increase the resilience of supply networks and ultimately facilitate digitalization. 
}

\onecolumn \maketitle \normalsize \setcounter{footnote}{0} \vfill
\input{Body/1_Introduction}
\input{Body/2_LRandbackground}
\input{Body/3_methodology}
\input{Body/4_disruption}
\input{Body/5_recovery}

\input{Body/6_evaluation}
\input{Body/7_conclusion}

\bibliographystyle{apalike}
{\small
\bibliography{example}}

\end{document}

%% file: Body/1_Introduction.tex
\section{\uppercase{Introduction}} \label{chap:introduction}
In highly globalized and complex \acp{sc}, performance analysis is essential as the change in behavior due to disruptive events does not only affect one organization but a highly connected network \cite{Singh2019}. 
The importance of systematic Disruption Management for Supply Chains was just recently again stressed in the course of the COVID-19 pandemic, but also already earlier in the light of events such as natural disasters, transportation blockages, sanctions etc.
Therefore, a vast share of enterprises rely on a Disruption Management Process (\ac{dmp}) to  monitor, model, assess and recover from  disruptions. 

The management and the evaluation of disruptions and their consequences on the \ac{sc} require the integration of various distributed data sources, e.g. from manufacturing, order and inventory management. 
\ac{sc} semantic models, i.e. ontologies, enable \ac{sc} data integration by providing a common and explicit understanding for business-related concepts ~\cite{pal2019integrating}.
Existing approaches address core \ac{dmp}  aspects but still in an isolated form, hence, limiting integrated \ac{sc} behavioral analysis.
Compared to previous work, our main contribution in this paper is MARE, MARE is a semantic disruption management and resilience evaluation framework, to integrate data covered by all \ac{dmp} steps \underline{M}onitor/Model, \underline{A}ssess, \underline{R}ecover and \underline{E}valuate.

MARE leverages a disruption ontology to model disruptive events and a knowledge-graph to represent specific disaster scenarios and the entailed effect on the \ac{sc}.
MARE includes production scheduling data and disruption knowledge-graphs to detect the implication of the disruption on the \ac{sc} operations, during the assessment phase.
Thus, MARE implements SPARQL-based recovery strategies to resolve the impairment caused by the disruption. 
Moreover, MARE incorporates a semantic evaluation framework to quantify the effect of recovery in terms of cost and delay on the \ac{sc}. 
Based on the evaluation results, and the recovery behavior analysis, \ac{sc} stakeholders potentially make decisions to redesign the \ac{sc} or establish new operational strategies ensuring a more resilient \ac{sc}. 

As a result, companies can rely on MARE  to integrate \ac{sc} data sources to  model and map the \ac{sc} behavior, to examine the effect of disruption and the consequences of applying various mitigation strategies.  
Ultimately, we deem that better simulation and analysis, as put forward by MARE, will contribute to mastering more complex \ac{sc} scenarios, control disruption accelerators e.g. the bullwhip effect and increase the resilience of supply networks. 
The remainder of the paper is divided as follows: first we introduce the background of the \ac{dmp} and existing semantic implementation for \ac{sc} disruption handling in Section~\ref{chap:motivation} and the motivation behind our proposed work. 
Second, we present MARE in Section~\ref{chap:methodology}, a framework to semantically model and manage disruptions and evaluate  \ac{sc} resilience.
In Section~\ref{chap:disruption}, we elaborate on MARE's semantic artifacts to model and assess disruptions, i.e, the first two stages of \ac{dmp}.
In Section~\ref{chap:recovery}, we introduce SPARQL-based recovery strategies to restore the \ac{sc} to the pre-disruption behavior. 
Also, we propose an evaluation framework to analyze recovery performance.
In Section~\ref{chap:evaluation}, we evaluate MARE to simulate the behavior of a synthetic \ac{sc} under various exemplary disrupted events.
Finally, we conclude and present an outlook for further steps to extend MARE in Section ~\ref{chap:conclusion}. 
\raggedbottom

%% file: Body/2_LRandbackground.tex
\section{BACKGROUND AND RELATED WORK} \label{chap:motivation}

\subsection{Supply Chain Disruption Management Process}
\ac{sc} disruptions as described by \cite{Craighead2007} are events that modify the flow of goods and materials, hindering the \ac{sc}'s overall objective to produce and deliver services and goods to end-customers.  
In fact, \cite{blackhurst2005empirically} define \ac{sc} \ac{dmp} as the process to discover the disruptive event, recover from the effect and potentially redesign the system triggered by recovery learning outcomes.  
Namely, discovery refers to the point in time when \ac{sc} stakeholders become aware of the disruption \cite{Macdonald2013}.
Then, disruption modeling of the system dynamics, e.g. via Petri nets, in simulation tools,  is essential in order to  analyze expected consequences and effects of the discovered event~\cite{bugert2018supply}. For instance, \cite{simulation} rely on the system dynamics simulation model  implemented in AnyLogic 8 tool \cite{tool} to demonstrate the behavior of a multi-echelon \ac{sc} responding to different end market scenarios. 

Further, \ac{sc} stakeholders choose the most effective recovery strategy to minimize the impacts of the disruption~\cite{Macdonald2013}.
Thus, the recovery performance analysis evaluates the \ac{sc} ability to repair and return to the pre-disruption phase.
Based on the evaluation's learning effects, \ac{sc} stakeholders can rethink the \ac{sc} design and operation processes and potentially decide on changes allowing more resilience e.g. increasing production capacity or applying a multiple sourcing strategy. 
The ability to both resist disruptions and recover the operational capability after disruptions occur, is defined as \ac{sc} Resilience~\cite{simbizi2021systematic}.

\ac{dmp} entails the integration of highly heterogeneous data sources. 
\cite{samaranayake2005conceptual}  elaborate that the integration provides visibility, flexibility and maintainability
of \ac{sc} components. 
Consequently, stakeholders can make more informative decisions towards enhancing \ac{sc} performance and increasing resilience. 
For instance, \cite{simchi2015identifying} integrate data from  bill of material, part routing, inventory levels, and plant volumes to map the \ac{sc} and accordingly assess the impact of a disruption  originating anywhere on product manufacturing and delivery. 
Also, \cite{sabouhi2018resilient} examine data from raw materials procurement along with inventory management systems to test the effect of various strategies in  establishing resilience.
\cite{ivanov2020digital} add that \ac{sc} digital twins enable integration to discover the link between \ac{sc} disruption and performance deterioration.
Namely, semantic models, one sort of digital twins, facilitate information exchange and allow \acp{sc} to reach full and agile information integration. 

\subsection{Semantic \ac{sc} Disruption Management}
In \cite{ascm} the authors explain that the view of \acp{sc} is based on internal data and seemingly relies on siloed or outdated data-sets.
Consequently, detecting emerging threats or calculating how disruption will unfold across the whole \acp{sc} and business units is generally possible but to a rather limited extent. 
However, semantic modeling of \acp{sc} allows to overcome the siloed paradigm and to blend and consolidate data from dispersed data sources \cite{ye2008ontology}.

There exist several articles in the literature that devise semantic implementations to analyze \ac{sc} performance during disruptions. 
\cite{Emmenegger2012} create an ontology model to monitor and model risks, give early warning and propose a procedure for assessing impacts on \ac{sc}. 
Also, \cite{palmer2018ontology} present an ontology-supported risk assessment approach for a resilient configuration of supply networks.
Moreover, \cite {Singh2019} provides an ontology-based decision support system to intensify the \ac{sc} resilience during a disruption. 
Despite these developments, we note that existing approaches address \ac{dmp} process steps in a rather isolated way, i.e., only one step of the process is incorporated e.g. to model the disruption risk or to assess its impact.  
Thus, we introduce MARE that, to the best of our knowledge, is the first work to integrate various data sources incorporated by all the \ac{dmp} steps to Monitor/Model, Assess, Recover and Evaluate. 

%% file: Body/3_methodology.tex
\section{METHODOLOGY} \label{chap:methodology}
\begin{figure}[tb]
  \centering
  \includegraphics[width=7.5cm]{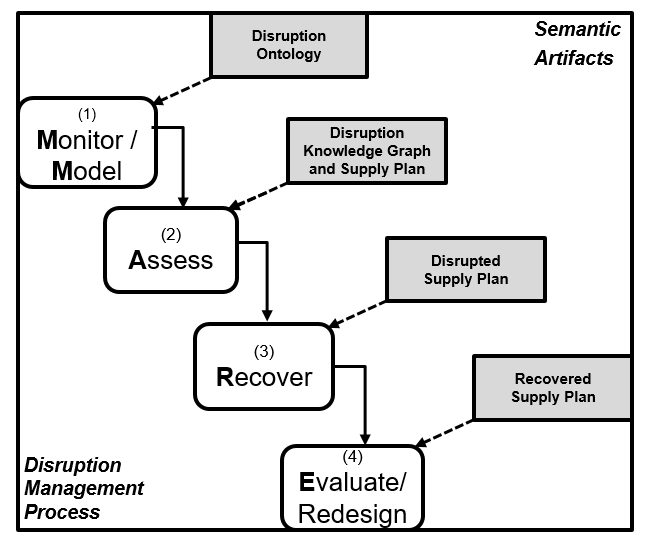}
  \caption{Overview in the MARE semantic disruption management and resilience evaluation framework.}
  \label{fig:methodology}
 \end{figure}

In this section, we describe our semantic disruption management and resilience evaluation framework, MARE. 
Moreover, we elaborate on MARE's semantic artifacts  i.e.,  ontologies, knowledge graphs and SPARQL to implement the \ac{dmp}.
As shown in \autoref{fig:methodology}, the \ac{dmp} starts with \textbf{Monitoring} and \textbf{Modeling} \ac{sc} disruptions. 
This phase is to discover the event that disrupts the \ac{sc} and to create a semantic model incorporating the disruption's attributes e.g. severity, cause and duration.
We rely on the \textit{Disruption Ontology} model, where the information is represented in the form of RDF triples\footnote{https://www.w3.org/TR/rdf-concepts/}, to establish a common understanding of a disruption event. 
Consequently, we create a specific instance of a disruption event i.e. \textit{Disruption \ac{kg}}.  
The output of the \textit{Monitor/Model} process step, the \textit{Disruption \ac{kg}}, is used in the following step to assess the effect of the disruption on the \ac{sc}. 

The target of a \ac{sc} is to fulfill end-customers' demand. 
Namely, \ac{sc} planning defines a scheduled capacity allocation for products among production facilities as well as the needed parts among suppliers i.e., \textit{Supply Plan}.
In previous work \cite{generator21}, we devised a semantic model for demand, production scheduling data and corresponding supply plan as follows: 
\begin{itemize}
    \item \textbf{Demand}: \ac{sc} demand is represented by the triples of the following form \textit{Customer makes Order}. 
    An order includes details about the product, delivery time and quantity:  \textit{Order hasProduct  Product},  \textit{Order hasDeliveryTime  xsd:dateTime} and \textit{Order hasQuantity  xsd:integer}.
    Based on the customer segmentation paradigm, customers are given a priority, entailing a certain sequence in demand fulfillment, i.e., \textit{Customer hasPriority xsd:integer}.

    \item \textbf{Supply Plan}: A supply plan is defined as the allocation of demand for parts among suppliers or the allocation of demand for products among production facilities~\cite{sawik2019multi}. \textit{Order hasSupplyPlan  Plan} and \textit{Plan needsPartner Partner} describe the needed \ac{sc} partners to fulfill this order. 
    Each partner is responsible for providing a product, i.e.  \textit{$<<$ Plan needsPartner Partner $>>$ getsProduct Product} at a certain time \textit{hasTimeStamp xsd:date}. The mentioned product can either be the final product or intermediary parts used to manufacture the final product. 
    The quantity and the price are modeled using \textit{hasQuantity xsd:double} and \textit{hasUnitPrice xsd:double}
\end{itemize}

Disrupted \ac{sc} partners potentially cannot fulfill their role in the plan, which affects the whole \ac{sc} performance.  
Therefore, during the disruption \textbf{Assessment} phase, we leverage queries adhering to the W3C SPARQL standard to identify affected \ac{sc} partners that are located in the same regions as the disruptions and who participate in the supply plan at the same time of the disruption (as described in detail in Section~\ref{chap:disruption}). 
In this process step, we integrate data sources from production scheduling (\textit{Supply Plan}) and disruption models (\textit{Disruption \ac{kg}}) to output the \textit{Disrupted Supply Plan}. 

The following step in the \ac{dmp} is to apply \textbf{Recovery} strategies to attempt a return to the pre-disruption performance of the \ac{sc}. 
In this phase, we rely on SPARQL endpoints to integrate data from production scheduling, order processing, inventory management, and suppliers assignment in order to find  alternative allocations for the disrupted plans.
The output of this step is one or more proposed \textit{Recovered Supply Plans} that include the updated scheduled allocations.

The last step of the \ac{dmp} is to \textit{Evaluate} the \ac{sc} recovery performance.
We propose a resilience \textbf{Evaluation} framework based on SPARQL queries to examine the time and the cost entailed by the \textit{Recovered Supply Plan} and required for the \ac{sc} to return to the pre-disruption state.
In fact, \ac{sc} stakeholders rely on this evaluation to potentially identify needs to redesign \ac{sc} or apply new operational strategies e.g. supplier diversification. 

%% file: Body/4_disruption.tex
\section{SUPPLY CHAIN DISRUPTION MODELING AND ASSESSMENT} \label{chap:disruption}
In this section, we present the first two steps of MARE to model and assess the effect of monitored disruptions on the \ac{sc}.  

\subsection{Modeling Disruption}

\subsubsection{Disruption Ontology}
\begin{figure}[!h]
  \centering
  \includegraphics[width=7.5cm]{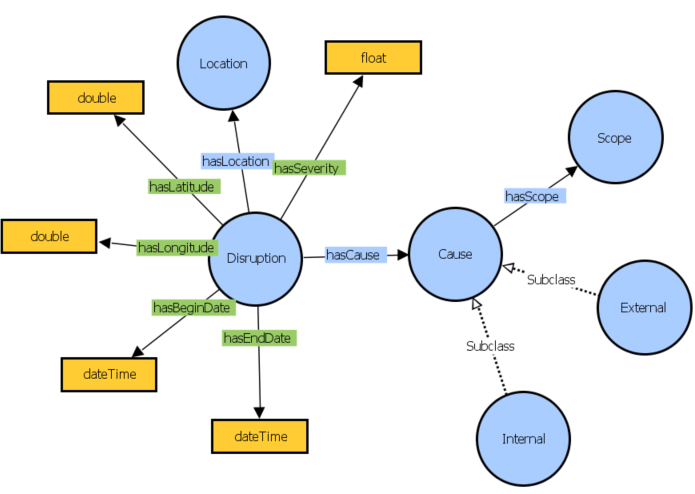}
  \caption{Overview on the core concepts of the Disruption Ontology for modeling disruptive event characteristics.}
  \label{fig:disruptionontology}
\end{figure}

We propose the \textit{Disruption Ontology} shown in  \autoref{fig:disruptionontology} to establish a model for disruptive events.  The ontology is based on RDF where the information is represented in triples. 
First, a triple of the following form \textit{Disruption hasCause Cause}, describes the main cause that led to the disruption. 
In fact, \cite{messina2020information} classifies disruption causes as internal and external. 
The first is caused by events happening within internal boundaries and the business control of the organizations e.g. malfunctioning of a machine or inventory corruption.  
While the latter is  driven by events either upstream or downstream in the \ac{sc} e.g. supplier insufficient capacity, interruptions to the flow of product, or significant increase/decrease in demand.  

Moreover, disruptions impact various \ac{sc} scopes e.g. production, logistics, inventory~\cite{Macdonald2013}. This, is reflected by triples of the form:  \textit{Cause hasScope xsd:string}. 
Additionally, the structure \textit{Disruption hasSeverity xsd:string}  incorporates financial losses caused by the disruption and their effect on the reduction or elimination of the production quantities. 
Further, disruption events can be of short or long duration. 
We use the following triple representation to model the disruption beginning and end \textit{Disruption hasBeginDate xsd:date} and \textit{Disruption hasEndDate xsd:date}. 
Also, we use \textit{Disruption hasLocation Location} to represent the geographical location where the disruption occurs. 
We rely on geo-coordinates system to resolve locations using the properties \textit{hasLongitude, hasLatitude}. 

In fact, classifying the modeled characteristics of the disruption enables \ac{sc} stakeholders to determine suitable recovery strategies for this event.
For example, in case of an external disruption due to the lack of a supplier's capacity, the recovery means can be to find an alternative supplier.
Whereas, to recover from an internal malfunctioning machinery within an own facility, one needs to fix it by retrieving spare parts from a machine of the same brand.

\subsubsection{Instantiated Examples}
The proposed disruption ontology incorporates disruption attributes to create a specific instantiation of a disruption event, represented by the \textit{Disruption \ac{kg}}.
We present in \autoref{tab:disruptions} various examples from past events to highlight possible variations in disruptions in terms of cause, scope, location, duration and severity.

\begin{table}[tb]
\caption{Disruption examples and corresponding triple representation.}\label{tab:disruptions} \centering
\includegraphics[width=7.5cm]{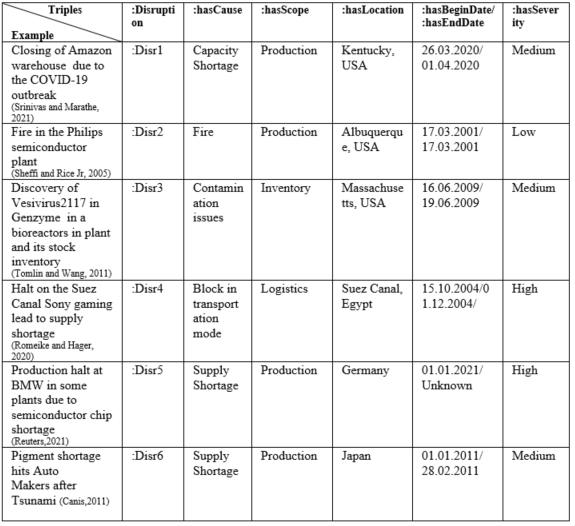}
\end{table}
 
\textit{:Disr1} is an example of capacity scarcity caused by labor shortage after a COVID-19 outbreak that led to a complete shutdown of production lasting four days.
\textit{:Disr2} shows a very short disruption, as the fire lasted for 10 minutes and the physical damages were minimal i.e., the severity is low. 
Further, the medium contamination described by \textit{:Disr3} affected not only the production plant but also the stockpile inventory. 

Moreover, due to a halt in maritime transportation mode caused by a blockage in the Suez Canal, Sony sales dropped from 70,000 a week to around 6,000, i.e. \textit{:Disr4}.
In fact, supply shortage includes scarcity in raw material or any event (bankruptcy, over-demand) that leads to a reduction or discontinuation in supply.
In 2020, due to the COVID-19 pandemic, automotive industry suffered from substantial drop in demand that led to slowing their semiconductor orders. 
Meanwhile, the semiconductor manufacturers faced a significant increase in demand due to the rising need for personal computers, servers, and equipment while their own facilities were shutting down because of COVID-19 outbreaks \cite{mckinsey}. 
For instance, \textit{:Disr5} representing over-demand, halted production and unstable orders, leads BMW to recognize a loss  of 30,000 units in production so far in 2021. This disruption has an undefined end date.
Similarly, \textit{:Disr6} models the missing color pigments produced by factories in Japan affected by the Tsunami in 2011. \textit{:Disr6} has medium severity since car manufacturers limit ordering vehicles only in specific shades. 

\subsection{Disruption Assessment and Effect}

After identifying and modeling the disruption, the following step is to assess the impact.
\ac{sc} disruptions potentially hurdle \ac{sc} entities from achieving operational goals i.e. fulfilling end customers orders.
We leverage data from production scheduling and order processing i.e. \textit{Supply Plan} along with the modeled disruption from the previous step i.e. the \textit{Disruption Knowledge Graph}. 

The first step to assess the disruption effect is to identify the \ac{sc} partners that are part of a supply plan, yet fall within the disruption location and time frame.
\autoref{lst:disruptedplan} \footnote{For simplicity, the query is just using a standard longitude/latitude matching, but in our implementation we actually implemented a geo-spacial rectangular containment matching between supplier and disruption locations.} retrieves and labels \ac{sc} partners and corresponding \textit{Disrupted Supply Plan}. 
Also, the effect of the disruption is defined by how many supply plans are affected. 
We insert  \textit{Disruption affectsPlan xsd:integer}  i.e. the count of disrupted plans identified in \autoref{lst:disruptedplan}.
\noindent\begin{minipage}{\columnwidth}
\begin{lstlisting} [caption={Identify Disrupted Partners.},label={lst:disruptedplan}]
INSERT { 
  ?plan :isDisrupted 'True' . 
  <<?plan :needsPartner ?partner>> 
    :isDisrupted 'True'.
  ?disruption :affectsPartner ?partner.} 
WHERE { 
  <<?plan     :needsPartner ?partner>> 
    :hasTimeStamp ?t.
  ?partner    :hasLongitude ?long . 
  ?partner    :hasLatitude  ?lat .
  ?disruption :hasLatitude  ?latitude .
  ?disruption :hasLongitude ?longitude.
  ?disruption :hasStartTime ?start .
  ?disruption :hasEndTime   ?end . 
FILTER (?t>=?start && ?t<?end &&
  ?longit=?long && ?lat=?latitude)
  }
\end{lstlisting}
 \end{minipage}

The second step is to size the effect of the disruption on the disrupted \ac{sc} partners. 
The severity of the disruption determines the impact of the event on the partner's capacity to fulfill the supply plan. 
For simplicity, we model the severity as a numerical factor that shows the reduction in production capacities caused by the disruption. 
As shown in \autoref{lst:disruptioneffect}, the pre-disruption allocated quantity is reduced by the severity factor. The difference between the original and the reduced quantities represents the quantity to be supplied or produced by alternative partners and means. 
\noindent\begin{minipage}{\columnwidth}
\begin{lstlisting} [caption={Determine Disruption Impact.},label={lst:disruptioneffect}]
SELECT * WHERE {  
  <<?plan :needsPartner ?partner>> 
    :isDisrupted   'True' ;
    :getsProduct   ?product ;
    :hasTimeStamp  ?t ;
    :hasQuantity   ?q .
  ?disruption :affectsPartner ?partner .
  ?disruption :hasSeverity    ?factor .
BIND (?q*?factor AS ?reduced) .
BIND (?q-?reduced AS ?toRecover)
}
\end{lstlisting}
 \end{minipage}
After modeling and assessing the disruption effect on the supply plans, the next steps in the \ac{dmp} are to implement recovery strategies and evaluate the \ac{sc} resilience and recovery performance. 

%% file: Body/5_recovery.tex
\section{RECOVERY AND RESILIENCE EVALUATION} \label{chap:recovery}
\subsection{Supply Chain Recovery}
In this section, we describe the implementation of the third step of MARE i.e., Recovery.
Recovery strategies are actions applied to regain the pre-disruption state of the \ac{sc}, capable of delivering products to customers on time while minimizing the cost. 
Via integrating data sources about inventory management, resources procurement, supply management and  logistics, we aim to recover disrupted supply plans. 
We present recovery strategies that rely only on the change in the \ac{sc} planning and do not require any physical modification in the industrial process as the latter are highly dependent on the industry. 
For instance, increasing production capacity or allowing faster production are not realistic in capital intense or complex industries like semiconductor production.
We propose the following SPARQL-based recovery strategies \footnote{We show only exemplary SPARQL queries for the recovery strategies and refer to the accompanying GitHub repository \cite{NRamzy2022} for the complete set.} capable of adapting the supply plan depending on the disruption cause and scope. For all the following queries we assume the recovery is for Product P, at time T in quantity Q.

\paragraph{S1: Strategic Stock} is defined as a stockpile of inventory that can be used to fulfill demand during a disruption~\cite{tomlin2011operational}. \autoref{lst:strategicstock} verifies if the partner has strategic stock and returns the required price. We use inventory management data sources to implement this strategy. In fact, storing the strategic stock entails costs for warehousing, labor and insurance.  
    \noindent\begin{minipage}{\columnwidth}
\begin{lstlisting} [caption={Strategic Stock  Strategy},label={lst:strategicstock}]
SELECT * WHERE {
:Partner :hasStartegicStock ?stock 
?stock :hasTimeStamp :T. 
?stock :hasQuantity ?q.
?stock :hasPrice ?price.
?stock :hasProduct :P.
FILTER (?q>= Q)
}
\end{lstlisting}
  \end{minipage}  
\paragraph{S2: Alternative Shipment} in case of a disruption affecting the transport mode e.g. flights, trains, a company can switch to another shipment mode to deliver products. The query in \autoref{lst:multipleshimpoent} retrieves the shipment modes employed by a partner and the entailed costs caused by the change of transportation modes, usually incorporated in logistics data sources \cite{messina2020information}. 
    
\begin{lstlisting} [caption={Alternative Shipment Recovery Strategy},label={lst:multipleshimpoent}]
 SELECT * WHERE {
 :Partner :hasTransportMode ?mode. 
 ?mode :hasCost ?cost.
 }
 \end{lstlisting}
 
\paragraph{S3: Delayed Recovery} this recovery strategy consists of verifying the status of the disrupted partner in case it can deliver slightly later than planned. \autoref{lst:delayed} checks for five days after the planned delivery time, if a \ac{sc} partner has enough capacity, lower than saturation, to fulfill the plan.  In fact,  small delays in deliveries can mitigate financial losses due to disruption \cite{paul2019managing}. Whereas, delays greater than five days (a week) potentially lead to fines of great amounts. Production management and scheduling data sources incorporate data about the continuous state of capacity production. 
   
  \noindent\begin{minipage}{\columnwidth}     
 \begin{lstlisting} [caption={Delayed Recovery Strategy},label={lst:delayed}]
 SELECT * WHERE { 
:Partner :hasCapacity ?cap.
:Partner :hasCapacitySaturation ?sat. 
?cap :hasProduct :P.
?cap :hasPrice ?price.
?cap :hasTimeStamp t_future.
?cap :hasQuantity ?q.
FILTER (?sat >= ?q + Q && t_future< T+5)
}
 \end{lstlisting}
 \end{minipage}

\paragraph{S4: Alternative Supplier} this strategy applies in case of an external disruption that hinders the supplier from providing the required products at the time included in the supply plan. 
In fact, \cite{sawik2019multi} elaborates that suppliers have production flexibility that allows them to deliver a contingency quantity in case other suppliers fail. However, the alternate source of supply can be more expensive than the firms’ primary suppliers, but it is deemed necessary,  in order to recover the disrupted supply plan~\cite{mackenzie2014modeling}.
To reduce purchasing prices and benefit from the high performance, suppliers that are capable of supplying the same products, are exchangeable \cite{hofstetter2019multi}. We model this via the property \textit{hasGroup}.
\autoref{lst:alternativesupplier} shows the query to find alternative, exchangeable suppliers that have the capacity (lower than saturation) to provide the same intermediate products or materials, for the same time as the disrupted supplier.  We rely on data from supply management and resources procurement to make decisions about suppliers belonging to the same group and their capacities.
\noindent\begin{minipage}{\columnwidth}
\begin{lstlisting} [caption={Alternative Supplier Recovery Strategy},label={lst:alternativesupplier}]
SELECT * WHERE{
:Partner :hasGroup ?group. 
?supplier :hasGroup ?group. 
?supplier :hasCapacity ?cap.
?cap :hasProduct ?p. 
?cap :hasQuantity ?q. 
?cap :hasPrice ?price. 
?cap :hasTimeStamp :T. 
?supplier :hasCapacitySaturation ?sat. 
FILTER  ( ?sat>= ?q + Q)  
}
\end{lstlisting}
   \end{minipage}


\paragraph{}
The output of this phase is a proposed \textit{Recovered Supply Plan} that minimizes recovery delays and costs. 
We identify a successful recovery as the case where all missing/reduced quantities from disrupted plans are provided alternatively. 
In this case, we insert the triple in the form \textit{Plan isRecoveredBy xsd:string}, where we explicit which recovery strategy applied. 

\subsection{Resilience Evaluation Framework}
In this section, we introduce step 4 in MARE  i.e., the evaluation framework for \ac{sc} resilience and recovery. 
Thus, we compare the pre-disruption supply plans to the recovered supply generated in the recovery phase.
We rely on the recovery performance evaluation metrics proposed by~\cite{Macdonald2013}. 

\paragraph{Recovery Cost Increase:} is the extra expense caused by the disruption and the recovery as compared to the original price of the pre-disruption supply plans.
First we  calculate the price of the recovered plan for each order and we retrieve the order original price. By summing the difference, we get the total cost increase for all orders in \autoref{lst:recoverycost}.
We do not consider the cost to rebuild anything lost physically as this is included in the severity factor. 
\noindent\begin{minipage}{\columnwidth}
\begin{lstlisting} [caption={Evaluate Recovery Cost Increase},label={lst:recoverycost}]
SELECT (SUM(?currentprice - 
?originalPrice) as ?costIncrease) {
SELECT ?originalPrice (SUM(?price) 
as ?currentprice 
WHERE {
  ?order  :hasPlan      ?plan;
  :hasOriginalPrice ?originalPrice. 
  <<?plan :needsPartner ?partner>>
    :hasQuantity   ?q; 
    :hasUnitPrice  ?p; 
    :hasTimeStamp  ?t.
  BIND (?p*q AS ?price)
} GROUP BY(?plan)
}

\end{lstlisting}
 \end{minipage}

\paragraph{Recovery Speed:} is the time taken till recovery is complete i.e., for \textbf{S3}, it is the next available day where there is enough production capacity, entailing a new delivery time. In \autoref{lst:recoveryspeed}, we calculate the number of orders where the delivery time in the supply plan is later than the original delivery time, pre-disruption. These orders are considered late orders, delayed by the difference between the original and the late delivery times.  

\noindent\begin{minipage}{\columnwidth}
\begin{lstlisting}[caption={Evaluate Recovery Speed},label={lst:recoveryspeed}]
SELECT 
  SUM(IF(?t>dt),1,0))  AS ?lateorders,
  SUM(IF(?t<=dt),1,0)) AS ?ontimeorders,
  SUM(?t-?dt) AS ?delay
WHERE {
  ?order  :hasPlan        ?plan. 
  ?order  :hasDeliverDate ?dt. 
  <<?plan :needsPartner   ?partner>>
    :hasTimeStamp  ?t
} 
\end{lstlisting}
\end{minipage}

\paragraph{Unsuccessful Recovery:} The ultimate goal of the \ac{sc} is to deliver finished products to end customers, yet the result of disruption caused by unplanned events can be unfulfilled orders as described by \cite{carvalho2012mapping}. This metric is the count of the supply plans where all missing/reduced quantities from disrupted plans are not provided alternatively i.e., \textit{Plan isRecoveredBy xsd:string} does not exist. 
This situation occurs in case there is no alternative shipment mode or there was no strategic stock available or if there were no substitute suppliers to supply alternatively. 
Moreover, when we apply \textbf{S3: Delayed Recovery} if there was no free capacity within the next five days, we consider this recovery unsuccessful. 

\paragraph{Customer Impact:} The previous metrics can be calculated by \ac{sc} stakeholders to analyze the impact of the disruption on specific customers. 
Within customer relationship management paradigm, \ac{sc} decision-makers apply recovery strategies in a way to attempt and reduce the impact of the disruption on high-priority customers.

%% file: Body/6_evaluation.tex
\section{EVALUATION AND DISCUSSION} 
\label{chap:evaluation}

In this section, we simulate the behavior of an exemplary \ac{sc} under various disruptions scenarios and evaluate the \ac{sc} recovery performance. 

\subsection{Experimental Setup}
The following details the experimental setup for the proposed evaluation. 

\textbf{Supply Chain Structure:} We consider a three-tier \ac{sc} network consisting of one central node, i.e, an OEM (Original Equipment Manufacturer) directly linked to four suppliers in supplier tier 1 and four customers in customer tier 1, where C1 is the customer with the highest priority. 

\textbf{Supply Chain Data:} We rely on the data generated and provided by the synthetic generator described in the technical report \cite{NRamzy2021}. 
We simulate 400 orders and their corresponding supply plans, generated for a time-frame of 178 days, i.e., half a year.
    
\textbf{Disruptions:} We simulate the disruptions listed in \autoref{tab:disruptions}. 
\textit{:Disr1-4} have internal causes, accordingly, we apply \textbf{S1: Strategic Stock, S2: Alternative Shipment, S3: Delayed Recovery} consecutively. 
While \textit{:Disr5} and \textit{:Disr6} are external, i.e., affecting suppliers, thus we apply \textbf{S4: Alternative Supplier}. 
Additionally, we create \textit{:Disr7,8} that occur internally and externally, thus we rely on a combination of the mentioned recovery strategies.
Moreover, for conciseness, we show \textit{hasDuration} which represents the length of the disruption in days, i.e, \textit{ hasEndDate} minus \textit{hasBeginDate}. 
The OEM in question relies on one transportation mode thus we cannot apply \textbf{S3: Alternative Shipment}. 

\subsection{Results}
\begin{table}[tb]
\caption{Resilience evaluation framework.}\label{tab:evaluation1} \centering
\includegraphics[width=7.5cm]{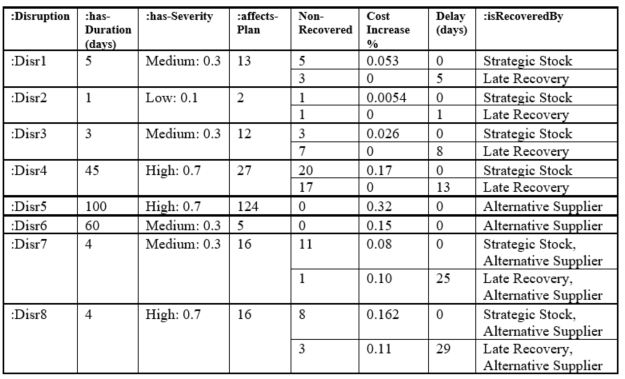}
\end{table}
We propose a resilience evaluation framework as shown in \autoref{tab:evaluation1} that incorporates the disruption characteristics i.e. duration, severity and the number of affected plans. 
Also, the framework includes the recovery metrics to evaluate the number of non-recovered plans i.e., unsuccessful recovery, the percentage of total cost increase and the delay. 
From the results in \autoref{tab:evaluation1}, we note that applying the strategic stock strategy leads to an increase in cost, whereas applying late recovery leads to delays in delivery.
This impact varies based on the duration and the severity of the disruption as well as the number of affected plans.
For instance, \textit{:Disr2} has a duration of one day and a low severity affecting only two plans, thus the cost increase and the delays entailed are minimal. However, \textit{:Dis1} and \textit{:Disr3} have medium severity and a duration of three and five days respectively, therefore, the cost and delay are more significant than in \textit{:Disr2}. 
Likewise, \textit{:Disr4} has a high severity and lasts for 45 days affecting 27 plans. Consequently the entailed cost and delay are higher than the previously mentioned disruptions. 
Also, we note that for \textit{:Disr5} and \textit{:Disr6}, there is a significant cost increase, since alternative suppliers can be more expensive than the firms’ primary suppliers. 

In case a disruption affects internally and externally \textit{:Disr7} and \textit{:Disr8}, there is a cost increase due to finding alternative suppliers and a delay in case of later recovery application. 
\cite{Macdonald2013} explain that the longer it takes to fully recover, the more expensive the entire recovery process is likely to be. 
The delays caused by \textit{:Disr8} are bigger than \textit{:Disr7}. Thus, the cost increase is greater as with high severity disruptions, the consequences are more severe. 

In order for stakeholders to make more informed decisions, they can rely on the customer impact analysis as shown in \autoref{tab:evaluation2} to examine the corresponding impact on specific customers. Consequently, they can decide which recovery strategy or combination of several to apply. 

\begin{table}[tb]
\caption{Customer impact evaluation with C1: customer with highest priority  .}\label{tab:evaluation2} \centering
\includegraphics[width=7.5cm]{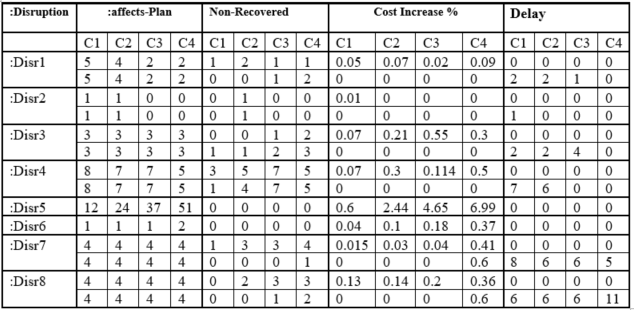}
\end{table}
It is important that while applying recovery strategies, orders made by customers with high priorities whose plans are disrupted, are recovered first. 
Therefore, we note that high-priority customers (C1) have fewer non-recovered plans. Therefore, their corresponding cost increase is higher than low-priority customers. Moreover, customers with low priority have longer delays because more important customers are recovered before, it might take more time periods to find the needed quantity to recover. 

\subsection{Impact and Discussion}
MARE is used to simulate the \ac{sc} behavior under various disruption scenarios. 
\ac{sc} stakeholders can make informed decisions based on the performance analysis to redesign into a more resilient \ac{sc} coping with unexpected events. We provide the following managerial insights: 

\begin{itemize}
    \item Behavior analysis, put forward by MARE, enables \ac{sc} stakeholders to decide on creating or modifying existing strategies. 
    In fact, some recovery strategies are only applicable in case pre-implementation approaches are established. 
    For instance, the OEM in the shown simulation did not support any alternative shipment mode, and consequently \textbf{S3} was not viable. 
    Similarly, a company can only apply \textbf{S4: Alternative Supplier} if the company has established a multiple sourcing strategy. 
    Also, the strategic stock recovery strategy requires the implementation of inventory management systems as well as replenishment. Likewise, decision-makers can decide to invest in extending the maximum capacity saturation to allow spare production capacity usable during disruption~\cite{chen2021supply}. 
    \item MARE supports supplier exchangeability, thus the cost increase caused by alternative suppliers can be reduced by establishing a wide \ac{sc} where suppliers are exchangeable. Consequently, the choice of an alternate source of supply is made easier in case of a disruption. 

    \item MARE provides \ac{sc} partners with knowledge about the impact of changes occurring in the production plan. Thus,
    MARE allows to reach full information integration to improve the selection of recovery strategies in future disruption occurrences.  Also, MARE enhances \ac{sc} visibility to mitigate the bullwhip effect. 

\end{itemize}

Nevertheless, MARE is limited as it only considers external disruptions that affect the supply. While sudden demand drops or surges can impact the \ac{sc} badly if the \ac{sc} is not equipped with suitable recovery strategies.  
Moreover, we focus only on recovery performance, whereas recovery structure and defining who from the \ac{sc} stakeholders is responsible and included in recovery, can potentially also be considered as explained by \cite{Macdonald2013}. 

%% file: Body/7_conclusion.tex
\section{CONCLUSION AND OUTLOOK} \label{chap:conclusion}

Recent events such as the COVID-19 pandemic, natural disasters, transportation blockages and political tensions resulting in sanctions have revealed the fragility of our highly globalized and complex \ac{sc} networks. 
Performance assessment for pre-disruption, during and post-disruption phases is needed to develop a resilient \ac{sc} network.
Namely, \ac{sc} integration, visibility and interoperability are essential for enriched \ac{sc} analysis to evaluate the behavior and facilitate decision making especially during irregular circumstances.
Semantic models enable \ac{sc} data integration and thus allow deep analysis while providing an overall perspective of the \ac{sc}. 
Existing semantic \ac{dmp} approaches address process steps in a rather isolated manner, i.e., only one step of the process is incorporated e.g. to model the disruption risk or to assess its impact.

With MARE we proposed a semantic disruption management and resilience evaluation framework, aligned with existing \ac{dmp} approaches, to integrate heterogeneous data sources (e.g. production scheduling, order processing), covered by all \ac{dmp} steps.
MARE relies on an ontology and \ac{kg} to \underline{M}onitor/Model a disruption. 
Then, MARE integrates data from production scheduling and order management to \underline{A}ssess, the effect of the disruption on the \ac{sc}. 
Next, MARE examines inventory management, procurement and suppliers assignment data sources to uncover various strategies to  \underline{R}ecover.  
The resilience framework is to \underline{E}valuate  the effect of the disruption on the \ac{sc} in terms of cost, delay and demand fulfillment.
Also, customer-specific metrics calculation allows to size the respective impact on customers. 

To ensure and enhance \ac{sc} resilience, \ac{sc} stakeholders can rely on the \ac{dmp} and resilience evaluation framework in MARE to extract decisions regarding \ac{sc} structure and operational strategies. 
MARE facilitates to grasp, control and ultimately enhance \ac{sc} behavior in complex \ac{sc} scenarios, tame disruption accelerators e.g. the bullwhip effect and increase the resilience of the supply network. 

The solid MARE framework being openly available on GitHub \cite{NRamzy2022} can be further extended to consider disruptions related to demand increase or drops and to examine combinations of recovery strategies in the comparison framework.
Also, MARE can be extended to include more recovery strategies e.g. spare capacity to check if the current utilization rate of the partner is below the saturation \cite{zsidisin2010perceptions}. 